# Critical behavior at de-pinning of a driven disordered vortex matter in 2H-NbS$_2$

Gorky Shaw[1], Pabitra Mandal[1], S. S. Banerjee[1,*], A. Niazi[2], A. K. Rastogi[3], A. K. Sood[4,+],

S. Ramakrishnan[5] and A. K. Grover[5,#]

[1]Department of Physics, Indian Institute of Technology, Kanpur-208016, India

[2]Department of Physics, Faculty of Natural Sciences, Jamia Millia Islamia University, New Delhi 110025, India

[3]School of Physical Sciences, Jawaharlal Nehru University, Delhi 110 067, India

[4]Department of Physics, Indian Institute of Science, Bengaluru 560012, India

[5]Department of Condensed Matter Physics and Materials Science, Tata Institute of Fundamental Research, Mumbai– 400005, India

*satyajit@iitk.ac.in

+asood@physics.iisc.ernet.in

#grover@tifr.res.in

Abstract

We report unusual jamming in driven ordered vortex flow in 2H-NbS$_2$. Reinitiating movement in these jammed vortices with a higher driving force, and halting it thereafter once again with a reduction in drive, unfolds a critical behavior centered around the de-pinning threshold via divergences in the lifetimes of transient states, validating the predictions of a recent simulation study, which also pointed out a correspondence between plastic de-pinning in vortex matter and the notion of *random organization* proposed in the context of sheared colloids undergoing diffusive motion.

Dynamical transitions in driven non-equilibrium systems have been considered to exhibit signatures similar to equilibrium critical phase transitions, e.g., the divergence in length or time scales with characteristic critical exponent[1,2,3,4]. Recent simulations and experiments on driven two-dimensional (2D)-colloidal systems have shown that the transition from a pinned state to a plastic flow state is associated with diverging time scales[5], along with a characteristic bimodal velocity distribution[6]. The initiation of sliding motion in vortex arrays in type-II superconductors is analogous to the generic *de-pinning transition* in condensed matter, such as, de-pinning of charge density waves, colloids, Wigner crystals, magnetic bubbles, magnetic-domain walls, etc[7]. A widely investigated non-equilibrium driven situation in superconducting vortex matter is the plastic de-pinning of vortices[7,8,9,10,11,12,13,14] wherein islands of pinned vortices coexist with filamentary channels of freely flowing vortices.

Application of an external transport current normal to the magnetic field gives rise to a Lorentz force which de-pins the vortex state to reveal a variety of different driven phases, e.g., a moving ordered Bragg glass state, the plastic flow state, etc.[12,13,14,15,16,17,18,19,20,21,22,23,24,25,26,27]. Using a new protocol, we have uncovered in a single crystal of a low $T_c$ anisotropic superconductor 2H-NbS$_2$, two limiting ends of the de-pinning force, viz., de-pinning of an ordered *pristine* pinned vortex state (having a lower critical current density, $J_c^l$) and that of a disordered vortex state (having a higher critical current density, $J_c^h$). In the present context, the disordered vortex state gets nucleated as the free flow (FF) in the driven ordered vortex state is *jammed* either by attempting to steadily accelerate or by waiting for long time. We also report on the discovery of vortex-velocity fluctuating between two extremes near the threshold limit of de-pinning the jammed, disordered vortex state. The life-time of the moving state in the fluctuating mode is found to diverge on approaching $J_c^h$ from above as well as below it, validating the prediction of a simulation study[5], which drew connection between plastic depinning and the phenomenon of random organization[4]. Few recent experimental studies[26,27] in periodically driven vortex matter pursuing reversible - irreversible transition[28] in periodically driven systems have also explored the above mentioned connection articulated in Ref. 5.

The isothermal electrical transport measurements have been performed at $T = 2.5$ K in the standard four probe geometry, with the dc magnetic field ($H$) applied parallel to the $c$-axis of the hexagonal single crystal of 2H-NbS$_2$, and with the dc current ($I$) flowing in its basal ($ab$) plane. The single crystals of 2H-NbS$_2$ were grown[29] by the standard vapour transport technique. The crystal used has dimensions of 0.9 x 0.9 x 0.045 mm$^3$, with $T_c$ of 5.8 K and the residual resistivity ratio, R(300 K)/R(10 K), of 35. When current $I$ sent thorough a superconductor exceeds the threshold de-pinning current ($I_c$), a voltage $V$ appears. We use a criterion of appearance of $V$ of 5 µV to determine $I_c$. The relationship $V=Hud$, where $d$ is the separation between the voltage contacts (~ 0.3 mm in the present case) helps in estimating the mean velocity ($u$), with which the vortices drift when $I > I_c$.

Figure 1 presents $V$-$H$ curves with $I = 15$ mA at 2.5 K, for two different field ramp rates, $dH/dt$ ($\dot{H}$). Note first that the de-pinning with a given $I$ (i.e., development of voltage) commences at a lower threshold field of 4 kOe for $\dot{H} = 300$ Oe/min, as compared to that near 13 kOe for $\dot{H} = 100$ Oe/min (corresponding to $u = 23$ cm/s)[30]. The $V$-$H$ response for $\dot{H} = 100$ Oe/min elucidates that $I_c(H)$ is greater than the applied $I$ of 15 mA for $H < 13$ kOe. After de-pinning at $H = 13$ kOe, the $V$-$H$ gradually enhances to reach its normal state value at upper critical field, $H_{c2}$. Considering that while ramping @ $\dot{H}$ =300 Oe/min, the de-pinning with $I$ = 15 mA commences at 4 kOe, the $I_c$ is > 15 mA only for $H$ < 4 kOe. Thereafter, as the $I_c(H)$ progressively decreases further (below 15 mA), the voltage (proportional[12] to a positive power of ($I - I_c(H)$)) continuously enhances from 4 kOe up to 10 kOe, after which the vortex flow is abruptly halted. This corresponds to a dynamic re-pinning, equivalent to a *jamming* transition in the driven vortex state. The *V-H* data in Fig. 1 establishes that the vortex states prepared with two different $\dot{H}$ have entirely different $I_c(H)$ values for 4 kOe < $H$ < 10 kOe. Above 10 kOe, the two *V-H* curves in Fig. 1 nearly overlap, and de-pinning commencing at about 13 kOe and the voltage response thereafter is independent of $\dot{H}$.

Figure 2 shows a set of *V-I* curves for vortex states at 8 kOe[31] prepared at 2.5 K via different $\dot{H}$, ranging from 100 Oe/min to 250 Oe/min. These curves clearly identify two distinct $I_c$ (at $H$ = 8 kOe) values, $I_c^l$ and $I_c^h$, as marked in Fig. 2. The inset (a) of Fig. 1 summarizes the variation of $I_c$ ($H$ = 8 kOe) with $\dot{H}$. The higher limiting value $I_c^h$ ($\approx$ 56 mA) is obtained with $\dot{H} \leq 180$ Oe/min, while the lower limit $I_c^l$ ($\approx$ 10 mA) identifies with $\dot{H} \geq 200$ Oe/min.

The *V-I* curve for the state prepared with $\dot{H}$ = 250 Oe/min in Fig. 2 has the usual inflection feature/knee shape[12,13] at $I = I_{cr}^l$, this *crossover* value[13] can be conveniently ascertained from *dV/dI* vs. *I* plot (cf. Fig. 14, Ref. [13]). Above $I_{cr}^l$, a quasi-linear *V-I* response sets in (implying steadily accelerating uniform flux flow (FF) regime), as *I* is gradually raised up to 88 mA. The state prepared with $\dot{H}$ = 200 Oe/min, also, has a low $I_c$ ($\approx$ 10 mA). However, in its quasi-linear flux flow regime above about 40 mA, it displays very large excursions, with *V* intermittently dropping to nominal zero, corresponding to the jamming of vortex flow. The fluctuations persist, and nearly cease only for $I > 75$ mA. The presence of fluctuations at $I > 60$ mA can also be witnessed in the *V-I* curves for the states prepared with $\dot{H} = 150$ Oe/min, where the initially prepared vortex matter has a higher critical current $I_c^h$. The higher *crossover* current limit, at which quasi-linear response sets in and large fluctuations in *V* largely cease, has been denoted by $I_{cr}^h$ ($\approx 83$ mA) in Fig. 2. For $\dot{H} = 100$ Oe/min, the *V-I* data are depicted for *I* ramped up to about 91 mA, and then back down to zero (indicated by the two oppositely directed short and thick arrows in Fig. 2). While ramping down, the *V-I* curve does not retrace its path below the *crossover* current $I_{cr}^h$. Larger excursions in *V-I* during ramp down commence as *I* is lowered below 56 mA ($= I_c^h$), however, they eventually cease, and the moving vortex state once again reorganizes into the disordered jammed vortex configuration, when $I \sim 30$ mA. On ramping up the current from zero value for the second time, the de-pinning commences once again only at the higher critical current $I_c^h$, and not at the lower critical current, $I_c^l$. This observation contrasts the finding in weakly pinned 2H-NbSe$_2$[14] that a vortex state with the lower de-pinning critical current is nucleated upon halting a steady flowing state created out of driving a super-cooled (*i.e., field cooled*) disordered vortex state with a higher $I_c(H)$.

The inset (b) in Fig. 1 summarizes the variation of $I_c^l$ and $I_c^h$ with *H* at 2.5 K. Due to collective interaction effects between the vortices, the $I_c(H)$ is known to depend inversely[14] on *H*. The field dependent lower $I_c^l(H)$ values are thus identified with the ordered vortex matter, and the nearly field independent (upto 10 kOe in the inset (b)) higher $I_c^h(H)$ values characterize the *disordered* counterparts[8,14,15] corresponding to independent pinning of small bundles of vortices.

The large fluctuations sampled in the vortex-velocity in Fig. 2 motivated us to record the time series of $V(t)$ response. Figure 3 collates $V(t)$ plots at $H = 8$ kOe (reached via $\dot{H} = 250$ Oe/min, after initial zero field cooling at 2.5 K) for six current values. In Figs. 3(a) to 3(c), the initial situation on the average (at $t = 0$) is a moving vortex matter (with finite voltage level, above 5 µV) and the eventual condition after a long waiting time ($\tau_l^j$) (ranging from ~$10^2$ to ~$10^3$ seconds) is a jammed state (i.e., $V \leq 2.5$ µV). When $I = 11.5$ mA (cf. Fig. 3(a)), the notionally ordered vortex matter having a lower critical current ($I_c^l \sim 10$ mA) is just depinned, and the *mean voltage level* along with the *average vortex velocity* is small (~ 2 cm/s), but, after wait time $\tau_l^j$ of about 1300 seconds, the $V$ abruptly drops to nominal zero level (i.e., < 5 µV). In Fig. 3(b), at $I = 28$ mA (> *crossover* current $I_{cr}^l$), the moving vortex state remains in quasi-linear flow mode, with small excursions anchored around $<V> \sim 150$ µV, for about 3000 seconds. However, at $I = 40$ mA (cf. Fig. 3(c)), $\tau_l^j$ is only about 350 seconds.

The $\tau_l^j$ values for $I_c^l < I < I_c^h$ are plotted in Fig. 4, and they display a non-monotonic behavior. The progressively faster moving vortex matter flow reorganizes into a disordered vortex configuration in a shorter time at $I > 30$ mA, with $\tau_l^j$ reducing to just about 50 seconds at $I = 50$ mA. We believe that enhancement in the driving current fuels a competition between the generation of defects[10,16,21], i.e., dislocations, and their subsequent annealing out in the driven vortex state[18]. One may therefore associate a net rate at which the topological defects, like dislocations, proliferate in the vortex state system. Therefore, $\tau_l^j$ is a measure of the drive ($I$) dependent time scale needed for a threshold density of defects to build up in the moving lattice at a given field in a given sample[32] beyond which the stability of the flowing vortex state is significantly compromised (which in the present case occurs beyond 30 mA), and the flow spontaneously reorganizes into disordered jammed vortex configuration. The vortex-motion induced crossover to a higher $I_c(H)$ state[33] reported earlier in YNi$_2$B$_2$C is notably different from the present result of a sudden change to a disordered state in 2H-NbS$_2$, due to similar field dependences of the higher $I_c(H)$ and lower $I_c(H)$ states in YNi$_2$B$_2$C. These two field dependent $I_c(H)$ states in YNi$_2$B$_2$C[33] are thus identified with ordered vortex lattices of different symmetry possible to observe in this system.

The disordered configuration realized after a wait time of $\tau_l^j$ can be de-pinned by enhancing the current beyond $I_c^h$ (= 56 mA). Figure 3(d) shows the $V(t)$ at $I$ = 60 mA. Note first that initially (i.e., up to about 50 seconds), the voltage level rapidly fluctuates between 20 µV and 60 µV and, thereafter, the upper limit of fluctuations enhances to reach up to about 190 µV. The 'state of fluctuations' lives up to about 1000 seconds (marked as $\tau_h^f$), and then suddenly a steady free flow (FF) (corresponding to $V$ ~ 160 µV in Fig. 3(d)) emerges. The $\tau_h^f$ measured for different $I$ ($> I_c^h$) are also plotted (as open circles) in Fig. 4. Note that $\tau_h^f$ values decrease rapidly as $I$ progressively increases above $I_c^h$. The identification of fluctuations and the divergence of $\tau_h^f$ as $I \rightarrow I_c^h$ in Fig. 4 are two of our key observations. The *notion* of *divergence* stands vividly illustrated in the $V(t)$ plot for $I$ = 56 mA in Fig. 3(e). At $I \approx I_c^h$, the large fluctuations in $V(t)$ range between just above zero level (closer to a pinned disordered state) to the high level of about 490 µV. Figure 3(f) shows that vortex matter initially driven into a steady flow with $I$ = 91 mA ($> I_c^h$), when slowed down by decreasing the current to 43 mA ($I < I_c^h$), maintains its flowing mode at the reduced current only for a short duration (~30 seconds). Thereafter, it fluctuates and eventually attains the disordered jammed vortex configuration. $\tau_h^f(I)$ measured for different $I$ ($< I_c^h$), reduced from initial current of 91 mA, have also been plotted (as filled circles) in Fig. 4.

Figure 4 shows that $\tau_h^f$ diverges from both below as well as above $I_c^h$, viz., the back and forth transformation between a steady flowing and a jammed vortex state is through the intervening fluctuating states, whose transient life times diverge upon approaching $I_c^h$ from either side of it[34]. The inset in Fig. 4 shows a fit of the $\tau_h^f(I)$ data to the relationship[5], $\tau \propto |(I - I_c^h)|^\beta$, with $\beta \approx$ - 1.60 (±0.12) for the disordered vortex matter at 8 kOe at 2.5 K[31]. The diverging timescales of large amplitude fluctuations ($\tau_h^f(I)$) in close vicinity of $I_c^h$ suggest that, like few other non-equilibrium systems, the FF vortex state can undergo a random organization into a pinned disordered configuration, whose de-pinning attribute is like a dynamical phase transition in a driven non-equilibrium system [2,3,4,5,25]. Recent studies suggest the parameter $\beta$, which is akin to a critical exponent[5], could imbibe the notion of universality class for dynamic transitions in diverse types of driven systems. In a different context[12,13], a case was made for the plastic depinning as a dynamical transition in 2H-NbSe$_2$ (see Fig. 25 in [13]). However, the present results of characteristic diverging time scales on both sides of the de-jamming threshold in 2H-NbS$_2$ establish the correspondence between the plastic depinning in the driven vortex state and a (dynamical) random organization occurring between a non-fluctuating (pinned or immobile) and a fluctuating (flowing) state in driven (diffusive) colloidal systems[4,5].

The bi-modal character of the moving phase fluctuating between two limiting values of the vortex velocities also echoes the behavior reported in the plastic flow regime of driven colloidal crystals[6], and in simulation studies[10] on driven vortex matter. Analysis of the probability distributions of vortex-velocities observed by us[35] shows that a fraction of the slower moving vortices coexist with a faster moving fraction. The analogue of vortex phase fluctuating between two extreme values of vortex velocities was not identified in the context of colloidal matter[6], which displayed the divergence characteristic in the transient times. The observation of negative values of vortex-velocities, opposite to the direction of drive (see, for instance, Ref. 35 and cf. Fig. 3) in the disordered jammed vortex state is also a new observation and justifies the 'jammed' nomenclature assigned to the disordered state in $2H-NbS_2$. Similar negative (entropy consuming) events have also been observed and analyzed through the non-equilibrium steady state fluctuation relation, in the sheared jammed state in surfactant-based hexagonal phase of cylindrical micelles[36].

Ulhas Vaidya is acknowledged for his help in experiments. S. S. Banerjee acknowledges the funding support from Department of Science and Technology of Govt. of India.

fluctuating state(s) present on either side of the $I_c^h$ has been reported prior to the present observations in 2H-NbS$_2$.

35. For the analysis of the probability distribution of vortex velocities in 2H-NbS$_2$, please see the website http://home.iitk.ac.in/~satyajit/Supplementary_info.pdf, and J.A.Drocco, C. J. Olson Reichhardt and C. Reichhardt, http://arxiv.org/abs/0909.2056v1 on negative fluctuations.

36. Sayantan Majumdar and A. K. Sood, Phys. Rev. Lett. **101**, 078301(2008).

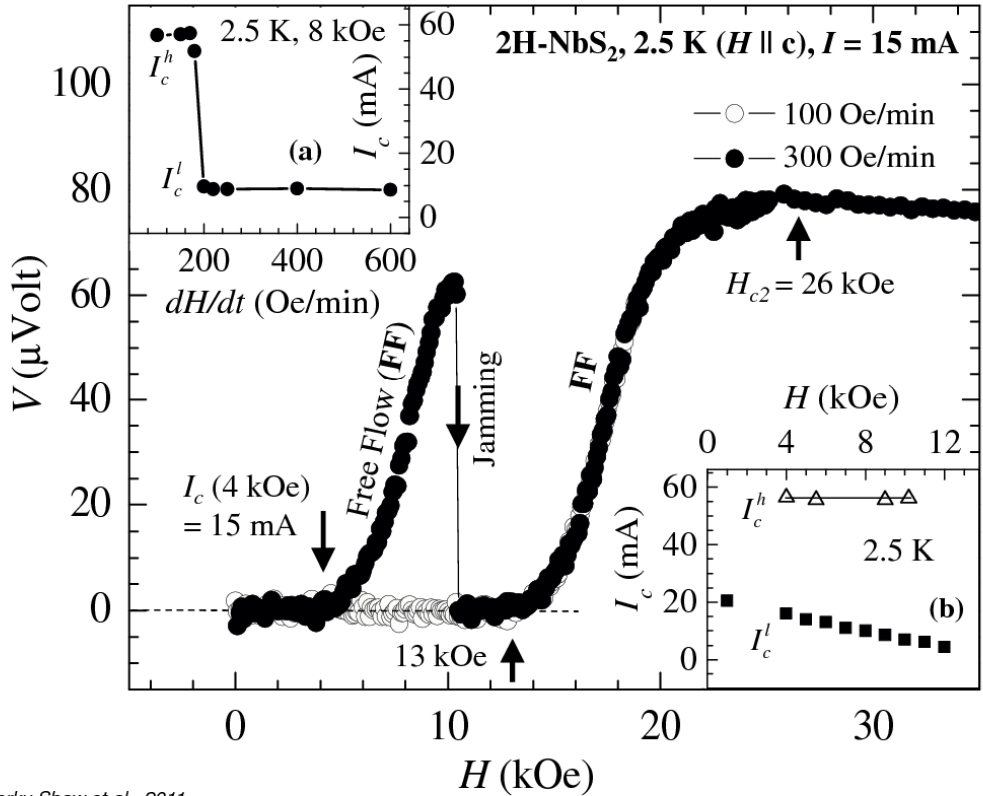

Fig. 1. *Gorky Shaw et al., 2011.*

Fig. 1: (Color online) Voltage drop versus field, *V-H*, across the single crystal of 2H-NbS$_2$ for $H \parallel c$ and at 2.5 K for $I = 15$ mA with $\dot{H} = 100$ Oe/min (open circles) and $\dot{H} = 300$ Oe/min (filled circles), respectively. The two arrows identify the threshold field values (13 kOe and 4 kOe, respectively) at which the *V(H)* rises above 5 μV. The *V(H)* for $\dot{H} = 300$ Oe/min displays a sudden transition from a high voltage (i.e., moving state) to the nominal zero value (i.e., jammed state) at $H = 10$ kOe ($\ll H_{c2}$). The inset (a) is a plot of $I_c$ vs. *dH/dt* at 8 kOe and the inset (b) displays variation of $I_c^l$ and $I_c^h$ vs. *H* (// c) at 2.5 K.

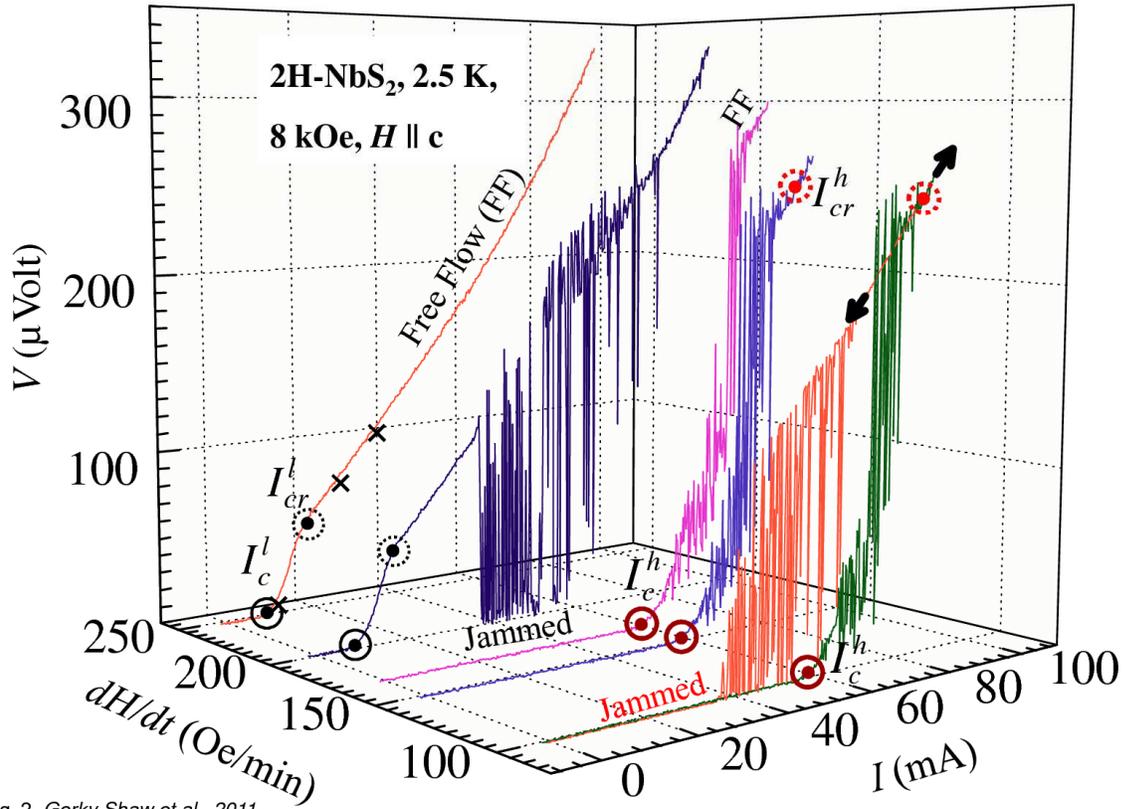

Fig. 2: (Color online) The *V-I* responses in 2H-NbS$_2$ for some selected vortex states at 8 kOe (*H* // c), prepared via different $\dot{H}$, ranging from 100 Oe/min to 250 Oe/min. The critical currents, $I_c^l$ and $I_c^h$, stand identified. The limiting values $I_{cr}^l$ and $I_{cr}^h$, above which *V-I* curves are quasi - linear are also marked. For vortex state generated via $\dot{H}$ = 100 Oe/min, a reverse leg of the *V-I* curve has also been shown. The cross marks (x) on the *V-I* curve for $\dot{H}$ = 250 Oe/min identify the three current values at which the time series measurements on the voltage are presented in Figs. 3(a) to 3(c).

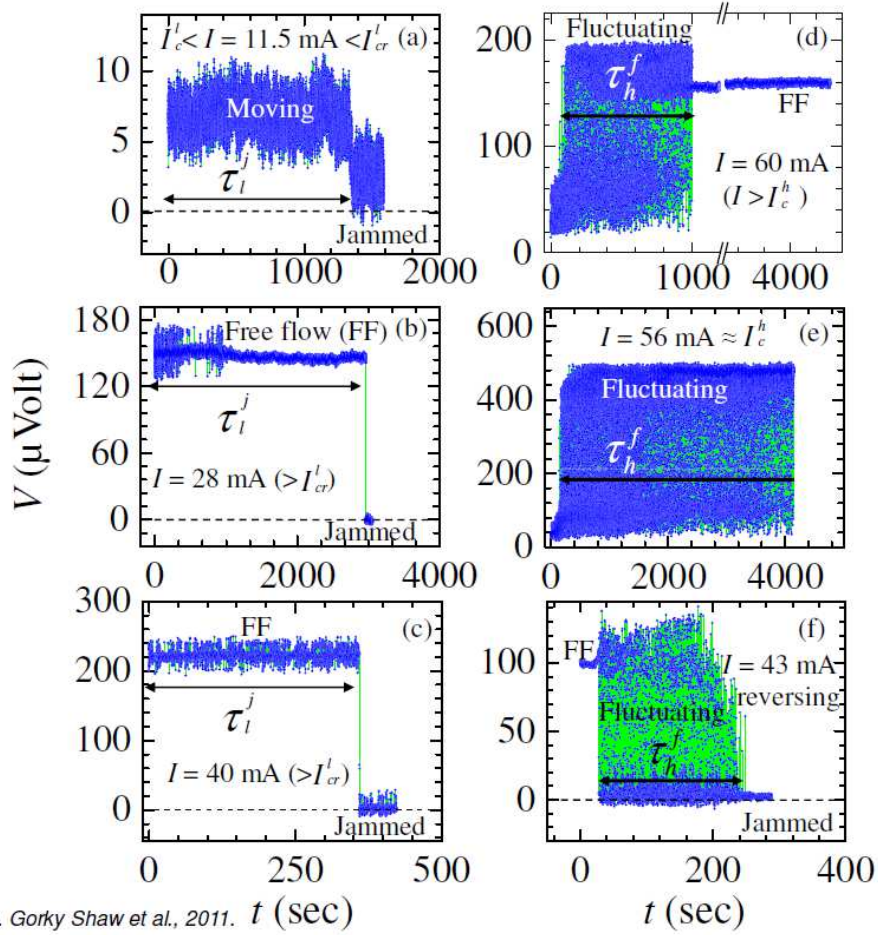

Fig. 3: (Color online) The panels (a) to (f) display time series measurements (data points shown as red open circles with the symbols joined by blue line) of voltage drop ($\propto$ vortex velocity) at different $I$ in a single crystal of 2H-NbS$_2$ at $H$ = 8 kOe (// c) and $T$ = 2.5 K, (a) $I$ = 11.5 mA ($> I_c^l$), (b) $I$ = 28 mA ($> I_{cr}^l$) and (c) $I$ = 40 mA, (d) $I$ = 60 mA, i.e., just after de-pinning the jammed/disordered state, (e) $I$ = 56 mA ($\approx I_c^h$), i.e., at the threshold of de-pinning the disordered state, and (f) $I$ = 43 mA, i.e., on reduction of current from free flow (FF) state at I= 91 mA to $I$ = 43 mA. The moving state ($<V>$ is greater than 5 µV) can be seen to transform into the jammed state ($<V> \sim$ 2.5 µV) after a lifetime of $\tau_l^j$ in panels (a) to (c). In panels (d) and (f) the fluctuating state can be seen to settle down to the free flow mode after a transient time of $\tau_h^f$. In panel (e), the fluctuating state seems to persist forever (*we recorded data up to about 4100 seconds*). Note also the overshoot of the vortex velocities below zero, i.e., to negative values in different panels.

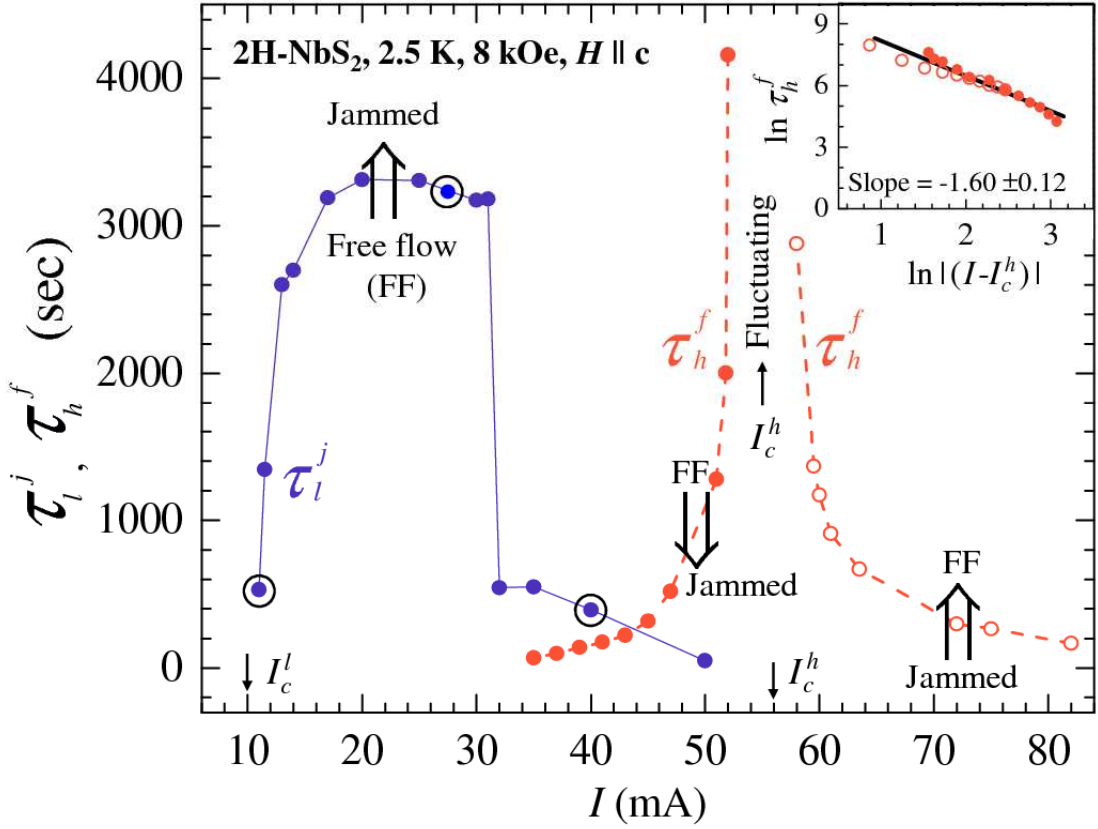

Fig. 4: (Color online) Variation of $\tau_l^j$ and $\tau_h^f$ with current $I$ for vortex states at 8 kOe (// c) and $T$ = 2.5 K. Three encircled data points on the $\tau_l^j$ ($I$) plot represent the typical $I$ values at which the time series data have been displayed in Figs. 3(a) to 3(c). An inset panel shows plots of log $\tau_h^f$ vs. log |($I$-$I_c^h$)| for two sets of data, for (i) $I > I_c^h$ (open circles) and (ii) $I < I_c^h$ (filled circles), respectively. The straight line drawn amounts to a power law relationship, $\tau_h^f \propto |(I-I_c^h)|^\beta$, with exponent, $\beta \approx$ -1.60 ± 0.12.